\documentclass[11pt,a4paper]{article}
\pdfoutput=1

\usepackage{jheppub}
\usepackage{dcolumn}
\usepackage{bm}
\usepackage{rotating}
\usepackage{verbatim}
\usepackage{xspace}
\usepackage{multirow}
\usepackage{arydshln}
\usepackage{amssymb}
\usepackage{slashed}
\newcommand{\bea}{\begin{eqnarray}}
\newcommand{\eea}{\end{eqnarray}}
\newcommand{\bem}{\begin{multline}}
\newcommand{\eem}{\end{multline}}

\newcommand{\beq}{\begin{equation}}
\newcommand{\eeq}{\end{equation}}
\newcommand{\be}{\begin{equation}}
\newcommand{\ee}{\end{equation}}
\newcommand{\bd}{\begin{displaymath}}
\newcommand{\ed}{\end{displaymath}}
\newcommand{\bc}{\begin{center}}
\newcommand{\ec}{\end{center}}
\newcommand{\ba}{\begin{array}}
\newcommand{\ea}{\end{array}}
\newcommand{\bt}{\begin{tabular}}
\newcommand{\et}{\end{tabular}}
\newcommand{\bi}{\begin{itemize}}
\newcommand{\ei}{\end{itemize}}
\newcommand{\bfig}{\begin{figure}}
\newcommand{\efig}{\end{figure}}
\newcommand{\sss}{\scriptscriptstyle}

\newcommand{\mcal}{\mathcal}
\newcommand{\esd}{\ensuremath{\epsilon_{sd}}\xspace}
\newcommand{\ecs}{\ensuremath{\epsilon_{\chi s}}\xspace}

\newcommand{\noast}{\vphantom{\ast}}
\newcommand{\nodag}{\vphantom{\dagger}}
\newcommand{\nobeta}{\vphantom{\beta}}
\newcommand{\tb}{\ensuremath{\tan\beta}\xspace}
\newcommand{\Bmumu}{\ensuremath{\mathrm{BR}(B_s \to \mu^+ \mu^-)}\xspace}
\newcommand{\Bsg}{\ensuremath{\mathrm{BR}(B \to X_s \gamma)}\xspace}
\newcommand{\Btn}{\ensuremath{\mathrm{BR}(B^+ \to\tau^+\nu_\tau)}\xspace}
\def\stilde{\widetilde}

\newcommand{\Neu}[1]{\ensuremath{\stilde \chi_{#1}^0}\xspace}

\newcommand{\doublet}[2]{\genfrac{(}{)}{0pt}{}{#1}{#2}}

\begin{document}

\title{Comparison of neutralino and sneutrino dark matter in a model with spontaneous CP violation}

\preprint{HIP-2012-21/TH, OSU-HEP-12-11}

\author[a]{Katri Huitu,}
\emailAdd{katri.huitu@helsinki.fi}
\affiliation[a]{Department of Physics, and Helsinki Institute of Physics,
  FIN-00014 University of Helsinki, Finland}

\author[b]{Jari Laamanen,}
\emailAdd{j.laamanen@science.ru.nl}
\affiliation[b]{Theoretical High Energy Physics,
  IMAPP, Faculty of Science,  Radboud University Nijmegen,
  Mailbox 79,
  P.O. Box 9010,
  NL-6500 GL Nijmegen, The Netherlands}

\author[a]{Lasse Leinonen,}
\emailAdd{lasse.leinonen@helsinki.fi}

\author[c]{Santosh Kumar Rai,}
\emailAdd{santosh.rai@okstate.edu}
\affiliation[c]{Department of Physics and Oklahoma Center for High
Energy Physics, Oklahoma State University, Stillwater, OK 74078, USA}

\author[a]{Timo R\"uppell}
\emailAdd{timo.ruppell@helsinki.fi}

\arxivnumber{1209.6302}

\abstract{%
Supersymmetric extensions to the standard model provide viable dark matter candidates and
can introduce additional charge-parity (CP) violation needed for obtaining the observed
baryon asymmetry of the universe. We study the possibilities of scalar and neutralino dark
matter with spontaneous CP violation in the next-to-minimal supersymmetric standard model
 with a right-handed neutrino. The observed relic density can be produced both by
a neutralino or a right-handed sneutrino as the lightest supersymmetric particle but when CP
is violated new annihilation channels become available and in general lower the relic density. We
consider collider phenomenology for a number of benchmark points which all satisfy experimental
constraints and have either the neutralino or the right-handed sneutrino
contribute to the dark matter abundance.
}



\maketitle

\section{Introduction}
\label{sec:intro}

One of the most compelling hints of physics beyond the standard model is the cosmological
observation that more than 80 \% of the mass of the universe is composed of dark matter
(DM) \cite{Komatsu:2010fb}. This evidence comes both from the astronomical observations of
gravitational effects at various scales and the cosmic microwave background measurements
which are consistent with such a large amount of dark matter.

Supersymmetry (SUSY) is one of the prime candidates of new physics and may provide an
appropriate dark matter particle. Supersymmetric models generally contain terms, which
violate baryon and lepton number conservation, and potentially lead to fast proton
decay. The presence of such terms can be prevented by conservation of $R$-parity
\cite{Salam:1974xa, Fayet:1974pd, Farrar:1978xj, Dimopoulos:1981dw, Farrar:1982te}. The
remarkable consequence of $R$-parity conservation is that the lightest supersymmetric
particle (LSP) will be absolutely stable. The LSPs left over after the Big Bang could then
explain the observed dark matter relic density (RD). The LSP is thought to be uncharged
(in both electric and color charges) so it interacts only weakly or through gravitational
interactions.

The candidates for the lightest supersymmetric particle in the minimal supersymmetric
extension of the standard model (MSSM) spectrum are the lightest neutralino, gravitino,
and sneutrino. The LEP-collider searches exclude a light left-handed sneutrinos as the
LSP, and masses beyond LEP's reach are ruled out by direct detection searches
\cite{Caldwell:1988su,Caldwell:1990tk,Reusser:1991ri,Mori:1993tj}. However, the
superpartner of the right-handed neutrino is a viable dark matter candidate
\cite{Asaka:2005cn,Asaka:2006fs}. A pure right-handed (RH) sneutrino has a very reduced
coupling to ordinary matter due to the smallness of the neutrino Yukawa coupling and is
therefore not produced thermally in the early universe.  If, however, the right-handed
sneutrino is by other means coupled to the rest of the observable sector it may be a
thermal relic and provide for an appropriate dark matter relic density.

Supersymmetry must be a broken symmetry at low energies. The explicit SUSY breaking is
introduced softly so that no quadratic divergences appear. This requires inclusion of all
the possible gauge invariant breaking terms in the Lagrangian. The large
number of breaking parameters parametrize the supersymmetry breaking, which is expected to
be spontaneous in a more complete theory. Soft SUSY breaking introduces a large number of
complex phases, which can lead to large CP violation effects \cite{Dugan:1984qf}. While
cosmological observations of the baryon asymmetry of the universe suggest that additional
CP violation beyond what the Standard Model (SM) offers is necessary \cite{Gavela:1994dt},
SUSY phases typically produce excessively large electric dipole moments (EDMs)
\cite{Ellis:1982tk,Polchinski:1983zd}. One way to manage this problem is to impose CP
conservation on the Lagrangian and introduce spontaneous CP violation (SCPV)
\cite{Lee:1973iz}. SCPV is an attractive solution since it radically decreases the number
of free CP-violating parameters, in addition to allowing study of the interplay between
different CP observables \cite{Hiller:2010ib}. It is well known that SCPV is not possible
in the MSSM and requires at least one and preferably two additional singlet fields
\cite{Masip:1998ui,Mohapatra:2011gp}. Therefore, we are led to consider SCPV in the
next-to-minimal standard model (NMSSM).

The NMSSM \cite{Ellis:1988er,Drees:1988fc,Ellwanger:2009dp} provides also a solution to
the so-called $\mu$ problem \cite{Kim:1983dt} by introducing an extra singlet scalar
superfield $\hat S$ (lepton number $L=0$), whose vacuum expectation value (vev) will
generate an effective $\mu$ term. This naturally leads to $\mu$ to be of the order of the
electroweak (EW) scale. The singlet also contributes to the Higgs masses already at the
tree-level, which may lead to heavier Higgses than in the MSSM. Adding yet another singlet
superfield $\hat N$ ($L=1$) to the NMSSM provides for right-handed neutrino and
sneutrino states. Since both $\hat S$ and $\hat N$ are gauge singlets it is tempting to
have one field to do the job of both. This, however, leads to fine tuning problems as well
as either explicit $R$-parity violation or spontaneous lepton number violation with a
superfluous Goldstone boson in the spectrum \cite{Goldstone:1962es}. Therefore, both
singlets are assumed to be included in the model. Interestingly, a non-vanishing vev of
the superfield $\hat S$ enables an effective Majorana mass term for the right-handed
neutrino similar to the effective $\mu$ term. Moreover, the presence of the singlet $\hat
S$ also leads to the electroweak scale interactions of the right-handed sneutrino with
other MSSM fields.

In this work we study neutralino and right-handed sneutrino dark matter in the framework
of an extended  NMSSM when CP is spontaneously violated. For the case of no CP violation,
the possible right-handed sneutrino dark matter in the NMSSM has been earlier investigated in
\cite{Cerdeno:2009dv,Cerdeno:2011qv} and neutralino dark matter in 
\cite{Cao:2011re,Vasquez:2012hn}. Here we first introduce the changes in the model due to 
the spontaneous CP violation. We scan over interesting ranges for a number of parameters. 
After taking into account all relevant experimental constraints it is seen that CP violation 
decreases the dark matter relic density because of new available decay channels. The 
possible collider signals are considered for several benchmark points, two of which have a
sneutrino as an LSP and five of which have a neutralino LSP.

\section{NMSSM with a singlet neutrino superfield}
\label{sec:model}

We start by introducing the important properties of the model. The superpotential for the
NMSSM with singlet superfields $\hat S$ (L=0) and $\hat N$ (L=1) is given by
\bea
\nonumber
W &=& \epsilon_{\alpha\beta}\left(
Y_E^{ij}  \hat H_1^\alpha \hat L_i^\beta \hat E_j^{\nobeta} +
Y_D^{ij}  \hat H_1^\alpha \hat Q_i^\beta \hat D_j^{\nobeta} +
Y_U^{ij}  \hat Q_i^\alpha \hat H_2^\beta \hat U_j^{\nobeta} +
Y_N^i \hat L_i^\alpha \hat H_2^\beta \hat N + 
\lambda \hat S \hat H_2^\alpha \hat H_1^\beta\right) \\
&&\qquad+\lambda_N \hat N \hat N \hat S +
\frac{\kappa}{3} \hat S \hat S \hat S,
\eea
where $\epsilon_{\alpha\beta}\,(\alpha,\beta =1,2)$ is a totally antisymmetric tensor with
$\epsilon_{12}=1$, $\hat X$ denotes a superfield and the rest are dimensionless couplings.
The soft SUSY breaking terms are
\bea
V_{\rm soft}&=&\Big\{\epsilon_{\alpha\beta}\left(
A_E^{ij} Y_E^{ij}  H_1^\alpha \stilde L_i^\beta \stilde E_j^{\nobeta} +
A_D^{ij} Y_D^{ij}  H_1^\alpha \stilde Q_i^\beta \stilde D_j^{\nobeta} +
A_U^{ij} Y_U^{ij}  \stilde Q_i^\alpha H_2^\beta \stilde U_j^{\nobeta} +
A_N^i Y_N^i  \stilde L_i^\alpha  H_2^\beta  \stilde N 
 \right. \nonumber \\
&& \qquad 
+ \left. A_{\lambda} \lambda  S  H_2^\alpha H_1^\beta\right) 
+ A_{\lambda_N} \lambda_N \stilde N \stilde N S +
\frac{A_{\kappa} \kappa}{3}SSS + \xi^3 S \Big\} + {\rm h.c.} \nonumber \\
&& \qquad
+ \, M_{\Phi,ij}^2 \Phi_i^\dagger \Phi_j^{\nodag} +
M_{\Theta,ij}^2 \Theta_i^{\noast} \Theta_j^\ast +
m_{H_1}^2 H_1^\dagger H_1^{\nodag} +
m_{H_2}^2 H_2^\dagger H_2^{\nodag} +
m_{S}^2 SS^\ast,
\eea
where $\Phi=\{\stilde L,\stilde Q\}, \Theta=\{\stilde E,\stilde N,\stilde U, \stilde D\}$
are the scalar components of the corresponding superfields. A discussion on the
theoretical merits of this and similar models (specifically the soft $S$ tadpole) can be
found in \cite{Frank:2005tn,Hugonie:2003yu,Ellwanger:2009dp}. For phenomenological merits
of this and similar models (NMSSM + right-handed neutrino) see
\cite{Abada:2010ym,Das:2010wp,Kang:2011wb}.

In this work we take only the third generation Yukawa couplings to be non-zero for
$Y_{U/D/E}$ and impose at the electroweak scale $Y_N^i=Y_N$ and $A_N^i=A_N$ for all three
generations. We also take all soft scalar masses diagonal, $M_{ij}^2=M^2\delta_{ij}$, and
take $M_U=M_D=M_Q$, as well as $M_E=M_L$. It is worth pointing out a tension in choosing 
natural sizes for some of
theses parameters. If we want the $\mu$ problem solved then $\langle S\rangle\sim$
EW scale and it follows that $M_N\sim \langle S\rangle\sim$ EW scale, assuming that
$\lambda$ and $\lambda_N$ are $\cal{O}$(1). Now the left-handed neutrino masses are
generated by the conventional seesaw mechanism:
\beq
M_{\nu}=\left(
\bt{cc}
0&$m_D$\\
$m_D^T$&$M_N$
\et\right)\to
m_{\nu_L}\simeq -m_D M_N^{-1}m_D^T,
\eeq
where $m_D\sim Y_N\times$ EW scale and thus $m_{\nu_L}\sim Y_N^2\times$ EW scale, meaning
we need to choose $Y_N\sim 10^{-6}$. Otherwise, even for $\langle S\rangle\sim M_N\sim
M_{Planck}$ we would have $Y_N\sim 10^{-2}$ on top of which we would have to choose
$\lambda\sim 10^{-17}$.

Spontaneous CP violation is introduced by complex vacuum expectation values of the $S$ and
$H_2^0$ fields,
\beq
\langle H_2\rangle = 
\doublet{0}{v_2 e^{i\delta_2}}, \qquad
\langle S\rangle = v_S e^{i\delta_S}.
\eeq
A complex phase of the vev of $H_1^0$ can be absorbed by field redefinitions ($\langle
H_1^0 \rangle = v_1$). Non-zero vevs for left- or right-handed sneutrinos would introduce
spontaneous R-parity breaking. Since we are interested in viable dark matter candidates we
consider only R-parity conserving models. Thus there are only two complex phases
$\delta_S$ and $\delta_2$ which enter couplings and mass matrices wherever the vevs of $S$
and $H_2^0$ appear. In particular the mass matrices of neutral scalars are no longer
block diagonal with respect to the division of CP even and CP odd gauge eigenstates. The
sneutrinos form an $8\times8$ mass matrix divided into $4\times4$ submatrices
\beq
M_{\tilde\nu}^2=\left(\begin{array}{cc}
m_{ee}^2&m_{eo}^2\\
m_{oe}^2&m_{oo}^2
\end{array}\right)
\eeq
where the subscript denotes CP-even/odd states. We introduce the following notation
\bea
A_i^{\pm\pm}&=&Y_N^i ( A_N^i v_2 \cos\delta_2 \pm 2\lambda_N v_2 v_S 
\cos (\delta_2 - \delta_S) \pm \lambda v_1 v_2 \cos\delta_S ),\\
B_i^{\pm\pm}&=&A_i^{\pm\pm}(\cos\to\sin),\\
\label{eq:Cpmpm}
C^{\pm\pm}&=&2\lambda_N A_{\lambda_N}v_S\cos\delta_S \pm 2\kappa\lambda_N v_s^2\cos2\delta_S 
\pm 2\lambda\lambda_N v_1 v_2 \cos\delta_2,\\
\label{eq:Dpmpm}
D^{\pm\pm}&=&C^{\pm\pm}(\cos\to\sin),\\
m_{L,ij}^2&=&M_{L,ij}^2+Y_N^iY_N^j(v_1^2+v_2^2)+\frac{1}{2}m_Z^2\cos2\beta\delta_{ij},\\
m_R^2&=&M_N^2+\sum_i {Y_N^i}^2(v_1^2+v_2^2)+4\lambda_N^2v_S^2.
\eea
The $4\times4$ submatrices are then
\bea
m_{ee}^2&=&\left(\begin{array}{cc}
m_{L,ij}^2&A_i^{+-}\\
A_j^{+-}&m_R^2+C^{+-}
\end{array}\right),\quad
m_{oo}^2=\left(\begin{array}{cc}
m_{L,ij}^2&A_i^{--}\\
A_j^{--}&m_R^2-C^{+-}
\end{array}\right),\\
m_{oe}^2&=&\left(\begin{array}{cc}
0_{3\times3}
&-B_i^{++}\\
B_j^{-+}&D^{-+}
\end{array}\right),\qquad\quad\,
m_{eo}^2=(m_{oe}^2)^T.
\eea
where $i,j$ are the family indices of the left-handed sneutrinos. The mixing between left-
and right-handed sneutrino states is suppressed by the smallness of neutrino Yukawa $Y_N$
in $A^{\pm\pm}$ and $B^{\pm\pm}$. The CP-violating mixing between even and odd states by
$B^{\pm\pm}$ can be further suppressed if the CP-violating phases are small. There is no
mixing between even and odd left-handed states and the mass splitting for left-handed
states is proportional to $Y_N^2$. Consequently our model has almost purely left-handed
mass degenerate sneutrinos which are CP eigenstates to a high degree of accuracy.

The dominantly right-handed sneutrinos are split by $C^{+-}$ and $D^{-+}$ into 
\beq
m_{\tilde\nu_{1,2}}^2\simeq m_R^2\mp\sqrt{{C^{+-}}^2+{D^{-+}}^2}.
\eeq
From Eqs.~\eqref{eq:Cpmpm} and \eqref{eq:Dpmpm} we see that the mass splitting is
proportional to $\lambda_N$ but  unsuppressed by possibly small CP phases due o the
complimentary nature of $C^{+-}$ and $D^{-+}$ with respect to the CP-violating phases. The
mixing between even and odd states is done by $D^{-+}$ which may be suppressed by small CP
violating phases.

The Higgs mass matrix is also no longer block diagonal but becomes a $6\times6$ matrix
where the CP-even and CP-odd states mix. The usual vacuum stability conditions are solved
for the EW scale mass parameters $m_{H_1}$, $m_{H_2}$, and $m_S$. The addition of CP phases
introduces new vacuum stability conditions which we solve by fixing the soft trilinear
terms $A_\lambda$ and $A_\kappa$. At tree level
\bea
\frac{\partial V}{\partial\delta_2}=0 &\to&
A_\lambda=-\kappa v_S \frac{\sin(\delta_2-2\delta_S)}{\sin(\delta_2+\delta_S)},\\
\frac{\partial V}{\partial\delta_S}=0 &\to&
A_\kappa=-\frac{3\kappa\lambda v_1 v_2 v_S \sin(\delta_2-2\delta_S)+\xi^3\sin\delta_S}{\kappa v_S^2\sin3\delta_S}.
\eea
In our numerical calculations we use the 1-loop effective scalar potential including the corrections
from the third generation (s)quarks to derive both the mass matrices and the vacuum
stability conditions.

The neutralino mass matrix is of the conventional NMSSM form with the addition of complex
phases
\beq
M_{\chi^0}=\left(\begin{array}{ccccc}
M_1&0&-\frac{g_1v_1}{\sqrt{2}}&\frac{g_1v_2}{\sqrt{2}}e^{-i\delta_2}&0\\
0&M_2&\frac{g_2v_1}{\sqrt{2}}&-\frac{g_2v_2}{\sqrt{2}}e^{-i\delta_2}&0\\
-\frac{g_1v_1}{\sqrt{2}}&\frac{g_2v_1}{\sqrt{2}}&0&-\lambda v_S e^{i\delta_S}&-\lambda v_2 e^{i\delta_2}\\
\frac{g_1v_2}{\sqrt{2}}e^{-i\delta_2}&-\frac{g_2v_2}{\sqrt{2}}e^{-i\delta_2}&-\lambda v_S e^{i\delta_S}&0&-\lambda v_1\\
0&0&-\lambda v_2 e^{i\delta_2}&-\lambda v_1&2\kappa v_S e^{i\delta_S}
\end{array}\right).
\eeq

\section{Constraints and the parameter space}
\label{sec:constrainstandparameters}

\subsection{Parameters}
\label{sec:parameters}

We carry out our numerical analysis of the effect of the CP-violating phases by randomly
sampling the parameters that affect the Higgs and sneutrino spectra as well as the
interactions of the dark matter candidate. The gaugino masses were left fixed to better
illustrate the effect of bino dominance in the neutralino LSP. The parameters affecting
squark masses were chosen such that we avoid the current experimental limits completely.
We have generated two data sets, one CP-conserving ($\delta_2=\delta_S=0$) and one where
we vary the CP phases along with the other sampled parameters. Table \ref{tab:parCommon}
gives the values for the fixed parameters and the ranges for the sampled parameters.

\begin{table}[ht]
\begin{minipage}[b]{0.49\linewidth}\centering
\bt{|c|c|}
\hline
$M_1$&300 GeV \\
$M_2$&600 GeV \\
$M_3$&1800 GeV \\\hline
$M_Q$&1000 GeV \\\hline
$A_t$&1500 GeV \\
$A_b$&1500 GeV \\
$A_\tau$&-2500 GeV \\\hline
$Y_{N_i}$&$10^{-6}$\\
$A_{N_i}$&0 GeV \\
\hline
\et
\end{minipage}
\begin{minipage}[b]{0.49\linewidth}\centering
\bt{|c|c|}
\hline
$\tb$&2 -- 50 \\
$\mu$&0 -- 500 GeV \\ \hline
$\lambda$&0 -- 0.8  \\
$\kappa$&0 -- 0.8  \\
$A_\lambda$&-1000 GeV -- 1000 GeV \\
$A_\kappa$&-1000 GeV -- 1000 GeV \\
$v_S$&$\mu/\lambda$\\ \hline
$\lambda_N$&0 -- 0.8  \\
$A_{\lambda_N}$&-1000 GeV -- 1000 GeV \\
$M_N$&0 -- 500 GeV \\
$M_{L,E}$&0 -- 500 GeV \\ \hline
$\delta_S$&$0-2\pi^\ast$ \\
$\delta_2$&$0-2\pi^\ast$  \\
$\xi$&-1000 GeV -- 1000 GeV \\
\hline
\et
\end{minipage}
\caption{\label{tab:parCommon} The parameters and sampling ranges used in our two data
sets. $^\ast$EDM considerations limit the range of the phases considerably and we use the
ranges $[0,0.3]$, $[\pi-0.3,\pi+0.1]$, and $[2\pi-0.1,2\pi]$ for generating points that
are subject to all of the experimental constraints.}
\end{table}

\subsection{Tools}
\label{sec:tools}

The computational tools we use are LanHEP \cite{Semenov:2010qt} for model file creation
and micrOmegas (v.2.4.1) for the main numerical analysis \cite{Belanger:2006is}.
Calculating the spectrum and diagonalizations are performed using the EISPACK
\cite{Garbow:1974sr} routines and calculation of electric dipole moments and rare decay
branching ratios is performed by a dedicated Mathematica code. B-physics constraints are
calculated by NMSSMtools \cite{Ellwanger:2005dv,Ellwanger:2006rn} using the supplied NMSSM
model.

There are two caveats to the use of NMSSMtools which we will briefly address here. One is
that the NMSSM model does not include a right-handed neutrino. However, loop
contributions to the calculation of \Bsg and \Bmumu that involve a right-handed sneutrino
will be suppressed by $Y_N^2/Y_{\tau/\mu}^2$ compared to the corresponding left-handed contributions and
are thus negligible in our model. The other caveat to using NMSSMtools is that CP
violation is not implemented in the supplied NMSSM model. Two factors can affect the calculation:
changes in the model's spectrum due to non-zero CP phases, and changes in the couplings of
the fields. For the particles appearing in the tree level and 1-loop contributions to
\Bsg, \Btn, and \Bmumu, the spectrum changes continuously from
$\delta_2,\delta_S=0\to\delta_2,\delta_S\neq 0$. To judge the impact of CP phases on the
calculations we did a comparative analysis using the MSSM for which both CP-conserving and
CP-violating code bases exist in NMSSMtools. In the limit of small CP phases we find that
the difference in the results for \Bsg, \Btn, and \Bmumu is at the 5\% level for phases as
large as $\pi/10$ and more generally at the level of 1\% for phases $<0.1$.

This gives us confidence that the CP-conserving NMSSM model can be used to calculate
$B$-physics constraints in the regime of small CP-violating phases ($\delta_2,
\delta_S<0.1$). Satisfying the EDM constraints restricts us naturally to this regime. We
find that in our CP-violating data set most points that satisfy EDM constraints have
$|\delta_2|<0.05$ (modulo $\pi$). The phase $\delta_S$ is not as constrained and in
general the phases $\delta_2$ and $\delta_S$ can both have large values such that the
physical phases entering EDM calculations are fine tuned to be small. However, to simplify
calculations and to be able to use the NMSSMtools package mentioned above we restrict both
phases to the ranges $[0,0.3]$, $[\pi-0.3,\pi+0.1]$, and $[2\pi-0.1,2\pi]$ when generating
points that are subjected to all the experimental constraints we impose.

\subsection{Constraints}
\label{sec:constraints}

In order to examine only physically relevant parameter regions, a number of constraints
are taken into account during our computations. We check the scalar sector for vacuum
stability and the sparticle spectrum against the most recent PDG limits
\cite{Nakamura:2010zzi}.

The LEP experiment places the lower limit on the standard model Higgs boson mass at 114
GeV. As the Higgs boson can only be detected by its decay products, not directly, a model
independent way to use the experimental results is to investigate the bounds on Higgs
boson couplings to $ZZ$ and $hZ$. In the NMSSM the couplings $hZZ$ and $hhZ$ may differ
from the standard model and consequently the mass of the Higgs boson can be below 114 GeV,
provided that the couplings are sufficiently small. The reduced Z couplings are
\bea
g_{h_iZZ}&=& \cos\beta\,\mcal{O}_{1i} + \sin\beta\,\mcal{O}_{2i}\\
g_{h_ih_jZ}&=&(\sin\beta\,\mcal{O}_{4i}+\cos\beta\,\mcal{O}_{5i})
(\cos\beta\,\mcal{O}_{2j}-\sin\beta\,\mcal{O}_{1j})
-(i\to j)
\eea
with $\mcal{O}_{ij}$ the 6 $\times$ 6 neutral Higgs mixing matrix. We use the most recent limits
these couplings impose on the Higgs mass, which are presented in \cite{Schael:2006cr}. The recent
results from the LHC and Tevatron experiments imply that the SM-like Higgs boson mass
should be around $m_H \sim 125$ GeV \cite{Collaboration:2012si,Chatrchyan:2012tx}. In our
analysis we thus also require one of the Higgs states to have a mass between 123 -- 128
GeV. The 125 GeV Higgs signal in the NMSSM has recently been studied in
Refs.~\cite{Ellwanger:2011aa,Gunion:2012zd,King:2012is,Vasquez2012qqq}.

The phases $\delta_2$ and $\delta_S$ that introduce CP violation in our model are flavor
diagonal and thus are expected to be highly constrained from, e.g, the electron electric
dipole moment measurements. Current limits \cite{Beringer:1900zz} constrain the
electron EDM to be below
\beq
d_e < 1.05\cdot 10^{-27} {\rm ecm} .
\eeq
The NMSSM specific contributions to \Bsg from the extended Higgs and neutralino
sectors arise only at two-loop level \cite{Ellwanger:2009dp} and, in general, are
considered to be small \cite{Hiller:2004ii,Domingo:2007dx}. In our calculations, the
theoretical uncertainties \cite{Misiak:2006zs,Ellis:2007fu} are combined with the present
experimental value \cite{Asner:2010qj}, which leads to \cite{Huitu:2011cp}
\begin{equation}
  \label{eq:4}
  BR(B \to X_s \gamma) = (355 \pm 142) \times 10^{-6}.
\end{equation}

The impact of the NMSSM on the branching ratio \Btn is only indirect, and the extra singlet
contributions are suppressed by the small neutrino Yukawas as mentioned earlier. The new
physics  contribution to the branching ratio can be quantified by defining a ratio
\cite{Isidori:2006pk,Bhattacherjee:2010ju}
\begin{equation}
  0.99<\frac{\mathrm{BR}(B^+\to\tau^+
    \nu_\tau) _{\mathrm{\,SM+SUSY}} }
  {\mathrm{BR}(B^+\to\tau^+\nu_\tau)_{\mathrm{SM}}}< 3.19,  
  \label{RatioExp}
\end{equation}
where the numerator denotes the branching ratio in the SUSY scenario. The constraint
\eqref{RatioExp} tends to prefer small values of \tb in order not to decrease the ratio
too much below the lower limit. A large charged Higgs mass decreases the new physics
contributions in general.

The effect of our model on \Bmumu at one loop comes from right-handed (s)neutrinos and is 
thus suppressed by the smallness of the neutrino Yukawa coupling. We apply the limit from \cite{Aaij:2012ac}
\beq
\Bmumu < 4.5 \cdot 10^{-9}
\eeq

The anomalous magnetic moment of the muon has been measured quite precisely. However, there is
uncertainty in the reliability of the theoretical prediction due to hadronic and
non-perturbative effects.  Therefore, we do not use the magnetic moment as a constraint.

The relic density of cold dark matter in the universe is determined by the Wilkinson
Microwave Anisotropy Probe (WMAP) \cite{Komatsu:2010fb} to be $\Omega_c h^2 = 0.1126 \pm
0.0036$.  If 10 \% theoretical uncertainty is added \cite{Baro:2007em} we find the
preferred WMAP range
\begin{equation}
  \label{wmaplimits}
  0.0941
  < \Omega_c h^2 <
  0.1311
\end{equation}
at  $2~\sigma$ level. As the dark matter may also contain a component of a
non-supersymmetric origin, we have used only the upper bound as a constraint.

\section{The effects of SCPV on the relic density and other constraints}
\label{sec:cp-violation}

In this section we show the effects of CP violation on the relic density of the LSP. All
plots contain points from the CP-violating data set unless otherwise specified. All points
also satisfy PDG constraints on the mass spectrum and vacuum stability.

Spontaneous CP violation is known to change the spectrum of the scalar sector in a
discontinuous manner. For small values of the phases there appears a light state, $h_S$,
in the spectrum \cite{Georgi:1974au}. Experimental constraints on the Higgs boson mass
force this light state to be singlet dominated \cite{Schael:2006cr}. We define the doublet
content of $h_S$ as
\beq
\esd = \sum_{i={1,2,4,5}}\mcal{O}_{i\,2}^{\,2}
\eeq
with $\mcal{O}$ the neutral Higgs mixing matrix and the free index $i$ runs over the
elements corresponding to the lightest Higgs' gauge eigenstates separated into CP-even and
CP-odd parts, i.e. $\{\mathrm{Re}H_1^0, \mathrm{Re}H_2^0, \mathrm{Re}S, \mathrm{Im}H_1^0,
\mathrm{Im}H_2^0, \mathrm{Im}S\}$. We find that for our data points the value of \esd is
between 0.1 -- $10^{-5}$. The presence of this light singlet state has different effects
depending on what kind of LSP we have.

The addition of the singlet to the model also opens up the possibility for the neutralino
LSP to have a significant singlino admixture and thus behave differently from its usual
MSSM counterpart. We denote the singlino component of the neutralino LSP as
\beq
\ecs = |N_{51}|^2,
\eeq 
where $N$ is the neutralino mixing matrix. We find that already at values of 0.1 the
behavior of, e.g., the relic density of the neutralino depends significantly on the Lagrangian
parameters associated with the singlino.

The other masses in the spectrum are also affected by the CP phases, but in a continuous
way, meaning that for small values of the phases the spectrum is very close to the CP
conserving spectrum. This means that in calculations of observables, such as the relic
density or rare decay branching ratios, the dominant effect comes from the new light scalar
introduced by SCPV in our model.

We already discussed that for B physics constraints the effect of small CP-violating
phases is negligible. We additionally find that the appearance of $h_S$ has very little
effect since it doesn't appear in the calculations of \Bsg, \Btn, and \Bmumu until two
loop order. The charged Higgses contribute already at lower order but their masses change
continuously with the CP phases and thus the effect of CP violation remains limited in the
regime of small CP phases. However, in the case of the relic density there are new,
possibly dominant, annihilation channels that open up with the appearance of $h_S$
depending on what kind of LSP we have.

Figures \ref{fig:feynChi} and \ref{fig:feynSnu} show the leading coefficients for two
classes of direct annihilation channels, i.e., those with $f\bar f$, $f\in [q,l,\nu]$, and
those with $h_Sh_S$ final states in the case of a neutralino or a sneutrino LSP,
respectively. We show neither the diagrams for vector boson final states nor for the
various co-annihilation channels; the coefficients are all sub leading or the same as for
 the $f\bar f$ diagrams. In the numerical calculations we naturally take
all these channels into account. As we can show, this rough approximation provides a good
understanding of the parametric dependence of the relic density. We have abbreviated the
typical electroweak coupling $\frac{e}{2s_Wc_W}\equiv C\simeq 0.37$.

\bfig[ht] 
\bc
\bt{clcl}
\includegraphics[width=0.24\textwidth]{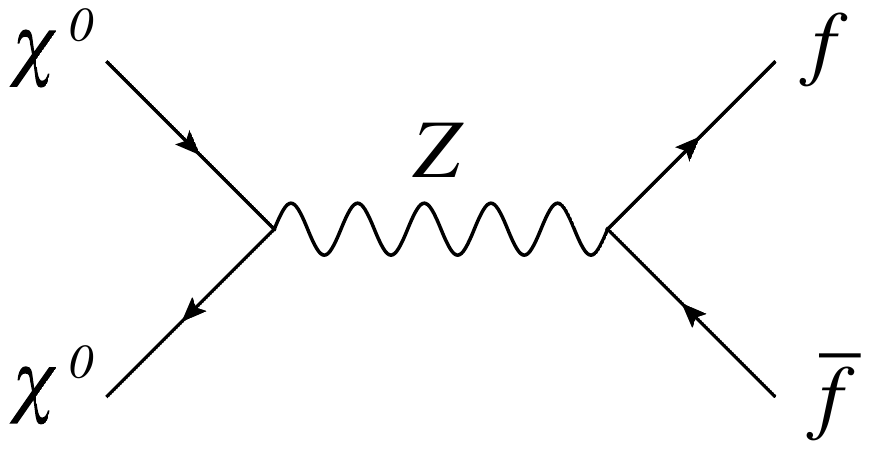}&
\raisebox{0.9cm}{
	\bt{l}
		$\buildrel d\over\sim C^2$\\
		$\buildrel s\over\sim 0$
	\et
	}&
\includegraphics[width=0.24\textwidth]{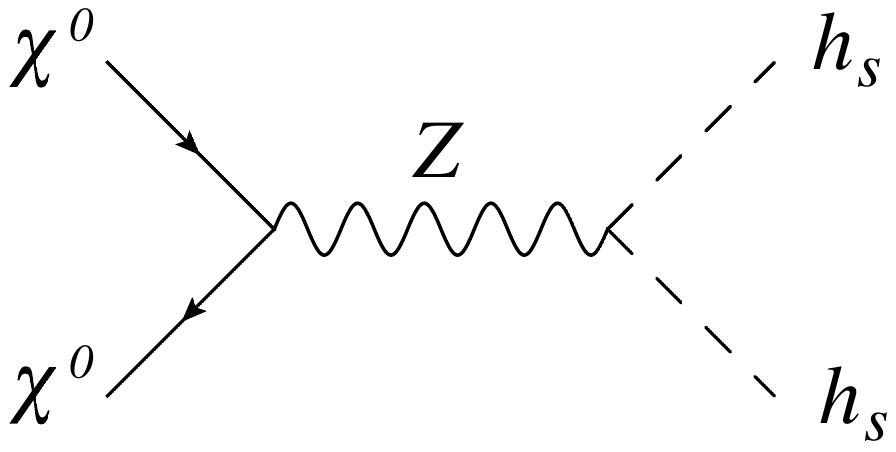}&
\raisebox{0.9cm}{
	\bt{l}
		$\buildrel d\over\sim C^2\esd^2$\\
		$\buildrel s\over\sim 0$
	\et
	}\\
\includegraphics[width=0.24\textwidth]{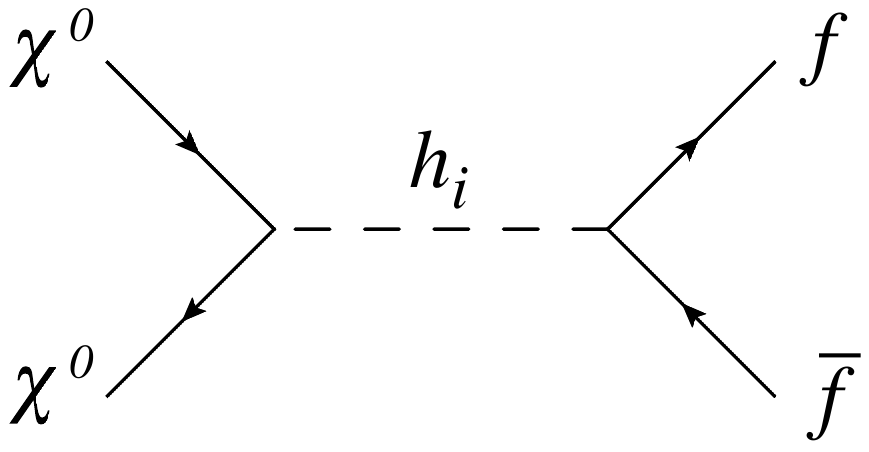}&
\raisebox{0.9cm}{
	\bt{l}
		$\buildrel d\over\sim C\cdot(Y_f,\lambda_N\esd)$\\
		$\buildrel s\over\sim \kappa\cdot(\lambda_N,Y_f\esd)$
	\et
	}&
\includegraphics[width=0.24\textwidth]{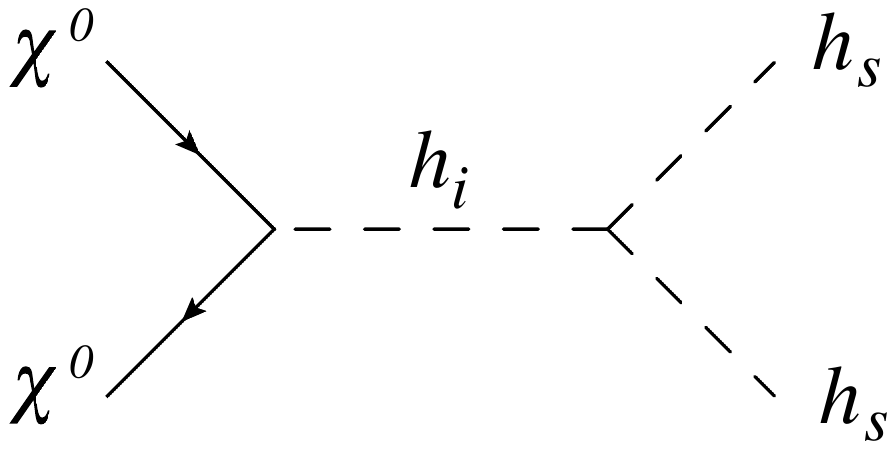}&
\raisebox{0.9cm}{
	\bt{l}
		$\buildrel d\over\sim C\esd\kappa$\\
		$\buildrel s\over\sim \kappa^2$
	\et
	}\\
\includegraphics[width=0.24\textwidth]{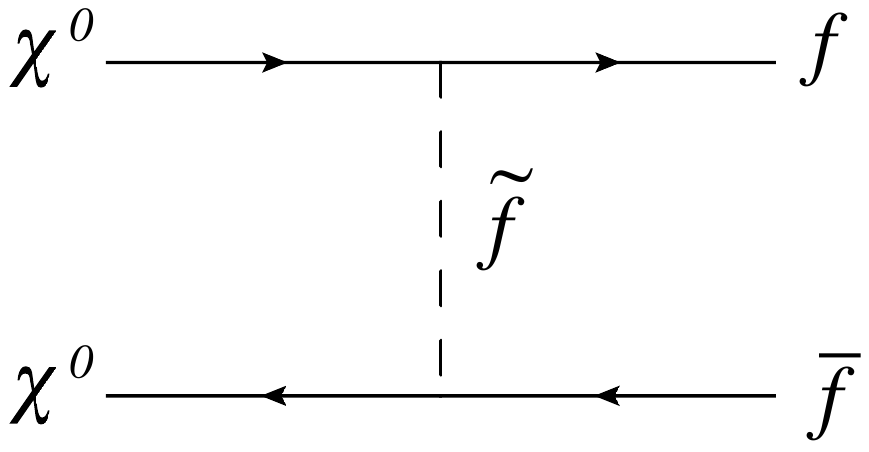}&
\raisebox{0.9cm}{
	\bt{l}
		$\buildrel d\over\sim (C,Y_f)^2$\\
		$\buildrel s\over\sim \lambda_N^2$
	\et
	}&
\includegraphics[width=0.24\textwidth]{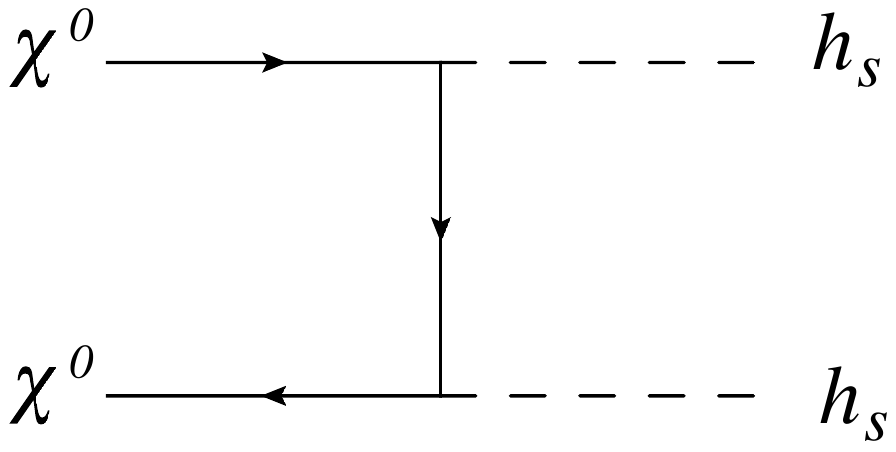}&
\raisebox{0.9cm}{	
	\bt{l}
		$\buildrel d\over\sim (\lambda,C\esd)^2$\\
		$\buildrel s\over\sim (\kappa,\lambda\esd)^2$
	\et
	}
\et
\caption{\label{fig:feynChi}Tree level pair annihilation channels for the neutralino into
$f\bar f$/$h_Sh_S$. Leading coefficients for the doublet (d) and singlet (s) component of
the neutralino are given. The addition of a light singlet scalar to the spectrum opens
additional channels.}
\ec
\efig
Fig.~\ref{fig:feynChi} shows the pair annihilation channels in the case of a neutralino
LSP. We have separated the leading coefficients for the case of a pure singlino as
``singlet" and group the couplings to all the other gauge eigenstates under ``doublet". We
see that for the doublet component of the neutralino the dominant contributions are $\sim
C^2$ with most of the new $h_S$ final states suppressed by \esd.

For the singlet component of the neutralino the story is different. The unsuppressed
$\sim\lambda_N$, $\lambda_N^2$ channels for the $f\bar f$ final states are available only
if the right-handed neutrino is lighter than the LSP. Even then, $m_{\nu_R}$ depends on
$\lambda_N$ directly so having $m_{\nu_R}$ small may also significantly suppress these
channels if, e.g., $\langle S \rangle$ is large. The channels for $h_S$ final states on
the other hand contain unsuppressed $\kappa^2$ dependence. However, $\kappa$ also controls
the neutralino mixing in a way that for small values of $\kappa$ the LSP becomes highly
singlino dominated. In that case, the only available annihilation channels are those now
suppressed by the smallness of $\kappa$, which leads to a larger relic density.

These effects are shown in Fig.~\ref{fig:RDvskappa}. In the left plot we can see that
for small values of $\kappa\lesssim 0.05$ none of the neutralinos have an appreciable
doublet component and also that the relic density for singlino dominated LSPs is controlled by
$\kappa$. In the right plot we see how with a growing singlino component the presence
of the new $h_S$ final state channels significantly decrease the relic density compared to
the CP-conserving scenario beginning at $\ecs\gtrsim 0.001$. We also see that once only
the singlino channels are available the relic density can again become very large due to
suppression by $\kappa$.
\bfig[ht]
\bc
\vspace{0.5cm}
\begin{tabular}{cccc}
\begin{sideways}\hspace{1.5cm}$\Omega_c h^2$\end{sideways}&
\includegraphics[width=0.4\textwidth]{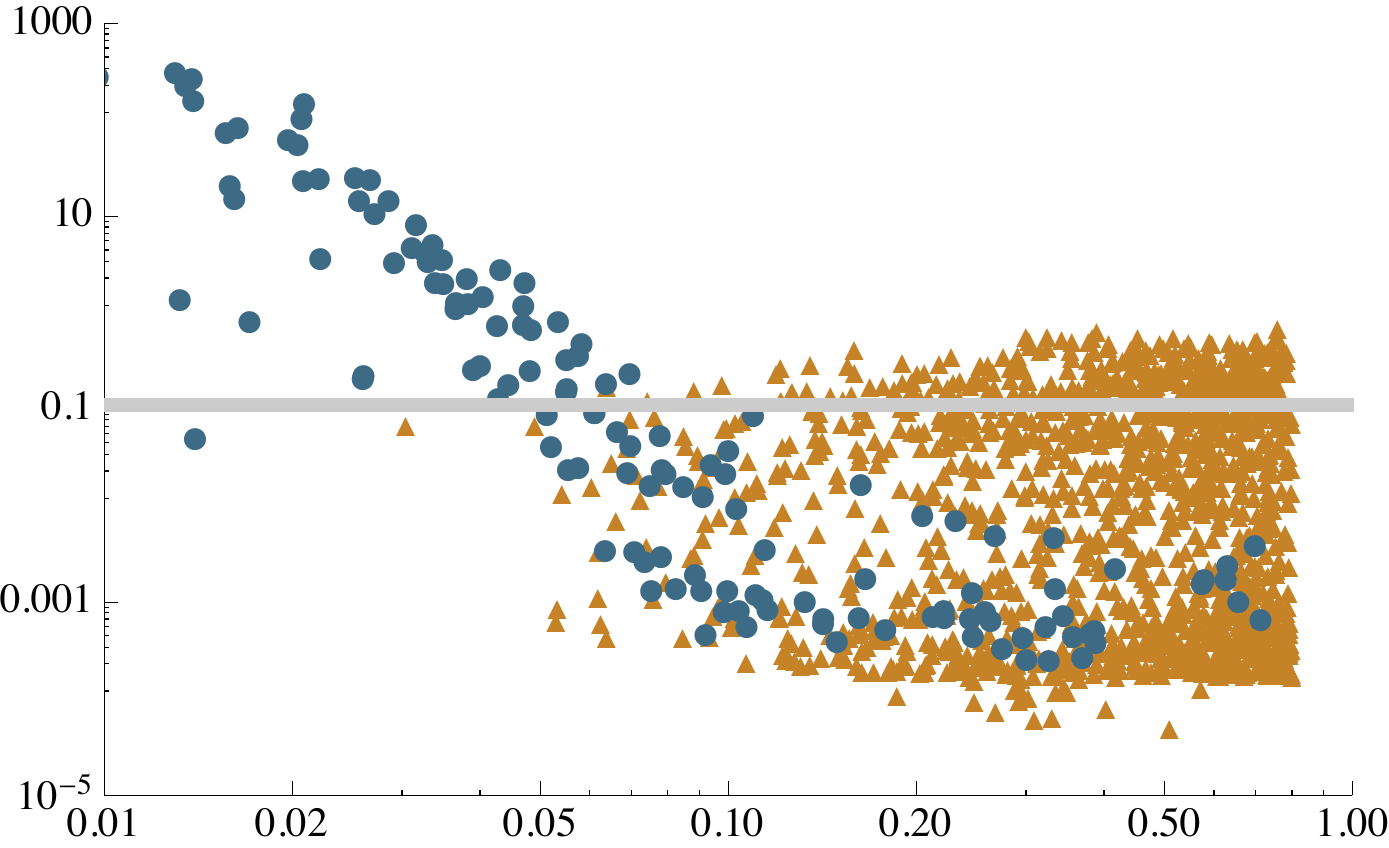}&
\begin{sideways}\hspace{1.5cm}$\Omega_c h^2$\end{sideways}&
\includegraphics[width=0.4\textwidth]{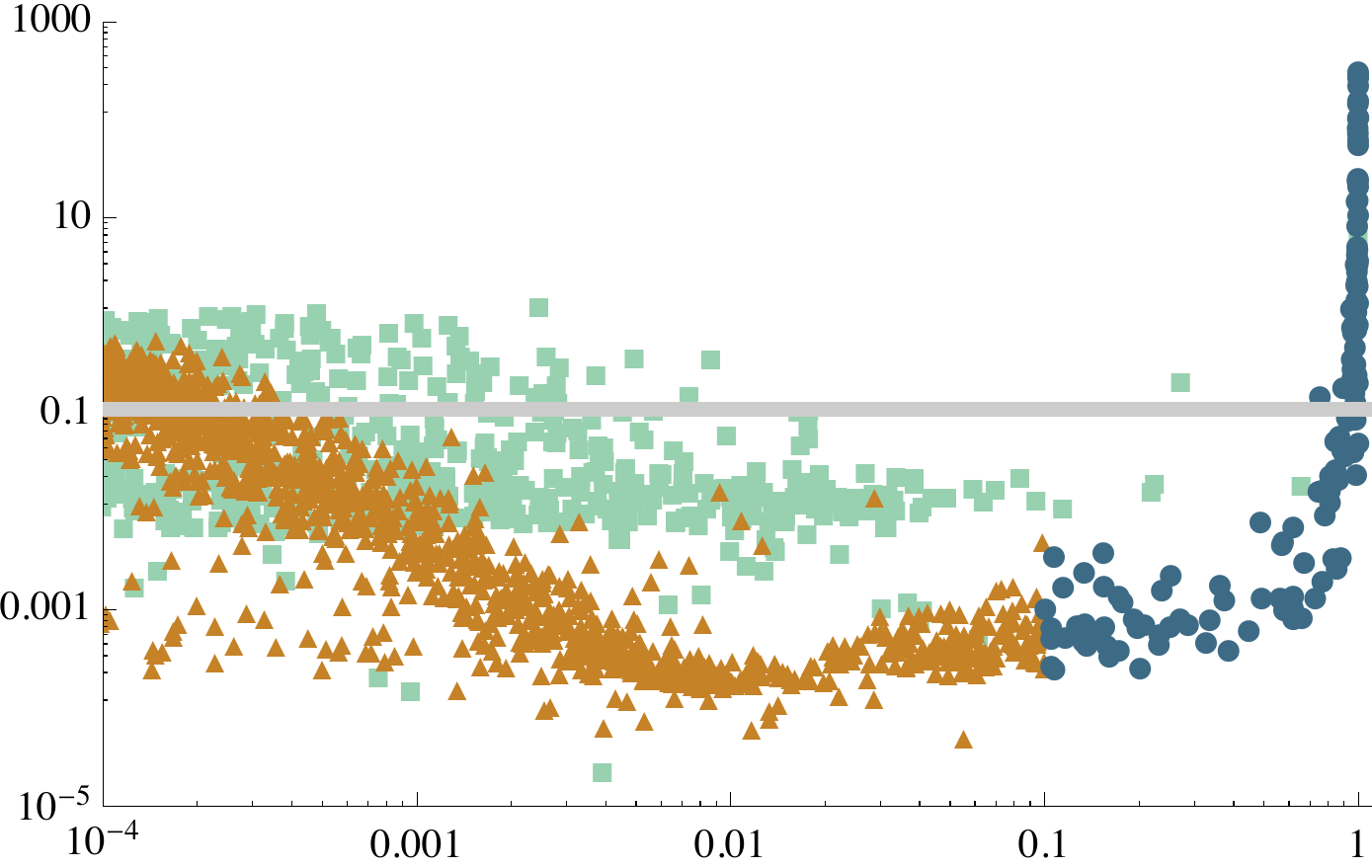}\\
& \raisebox{0.1cm}[-0.1cm]{$\kappa$}
&& \raisebox{0.1cm}[-0.1cm]{\ecs}
\end{tabular}
\vspace{-0.3cm}
\caption{\label{fig:RDvskappa}Left: The relic density against the trilinear coupling $\kappa$ 
for neutralino LSPs. Right: The relic
density against the singlino component of the neutralino LSP. Points with (blue circles) 
and without (orange triangles) significant singlino
component. Green boxes depict points from the CP-conserving data set. The grey band
indicates the the current WMAP limits on the relic density.}
\ec \efig

\bfig[ht] \bc
\bt{clcl}
\includegraphics[width=0.25\textwidth]{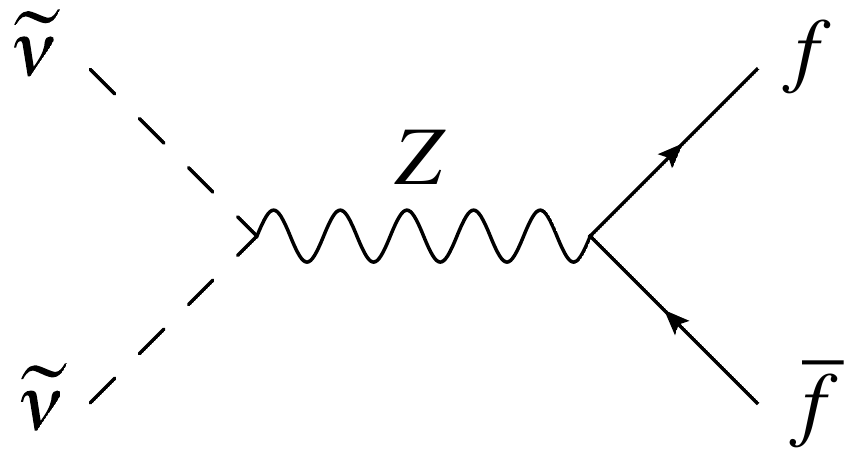}&
\raisebox{0.9cm}{	
	\bt{l}
		$\buildrel \sss{L}\over\sim C^2$\\
		$\buildrel \sss{R}\over\sim 0$
	\et
	}&
\includegraphics[width=0.25\textwidth]{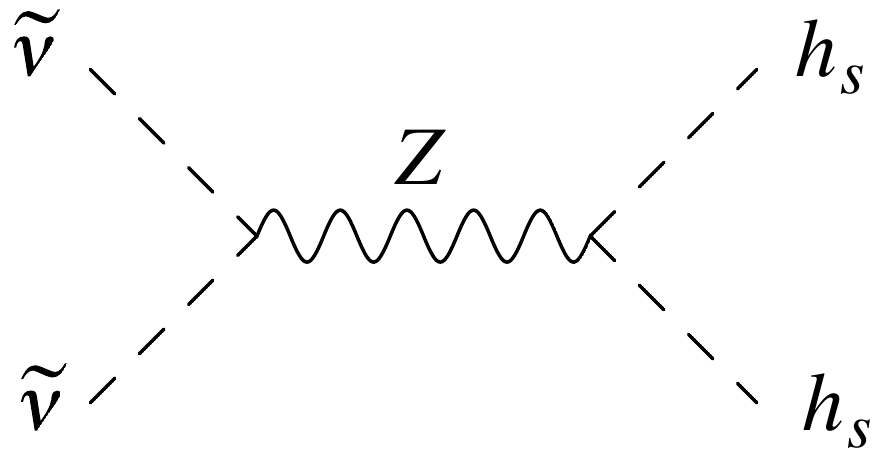}&
\raisebox{0.9cm}{	
	\bt{l}
		$\buildrel \sss{L}\over\sim C^2\esd^2$\\
		$\buildrel \sss{R}\over\sim 0$
	\et
	}\\
\includegraphics[width=0.25\textwidth]{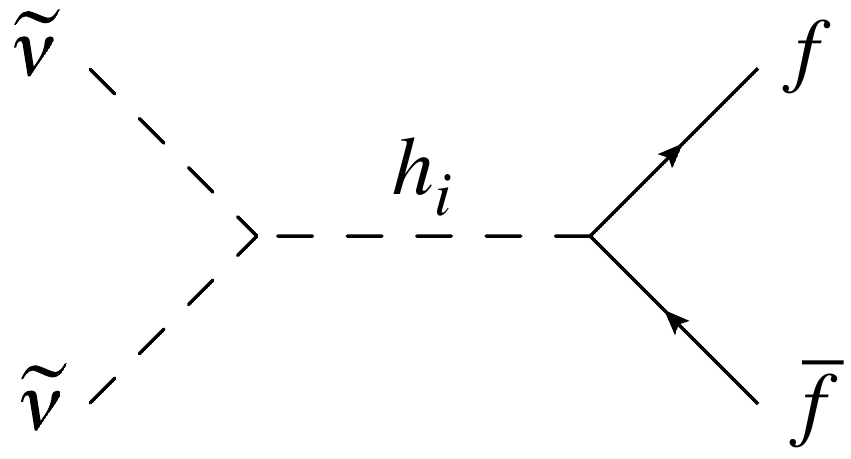}&
\raisebox{0.9cm}{	
	\bt{l}
		$\buildrel \sss{L}\over\sim C^2Y_f^2$\\
		$\buildrel \sss{R}\over\sim \lambda_N\kappa$
	\et
	}&
\includegraphics[width=0.25\textwidth]{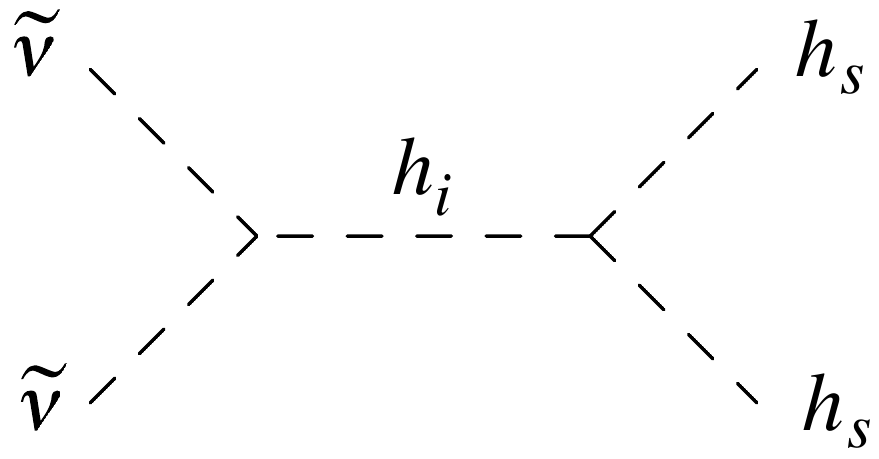}&
\raisebox{0.9cm}{	
	\bt{l}
		$\buildrel \sss{L}\over\sim C^4\esd^2$\\
		$\buildrel \sss{R}\over\sim \lambda_N\kappa$
	\et
	}\\
\includegraphics[width=0.25\textwidth]{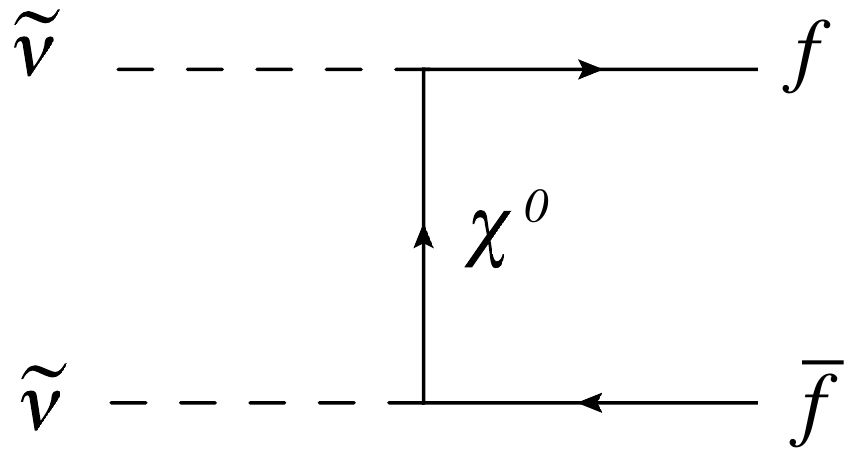}&
\raisebox{0.9cm}{	
	\bt{l}
		$\buildrel \sss{L}\over\sim C^4$\\
		$\buildrel \sss{R}\over\sim \lambda_N^2$
	\et
	}&
\includegraphics[width=0.25\textwidth]{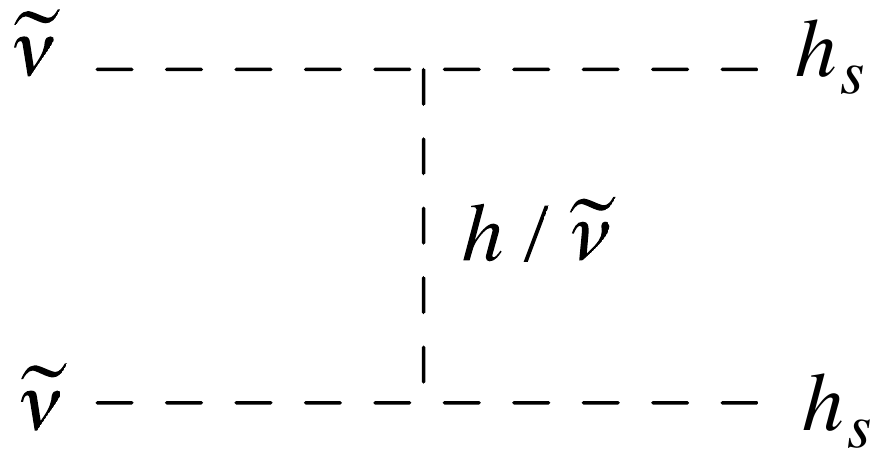}&
\raisebox{0.9cm}{	
	\bt{l}
		$\buildrel \sss{L}\over\sim C^4\epsilon_{sd}^2$\\
		$\buildrel \sss{R}\over\sim \lambda_N^2$
	\et
	}\\
&&\includegraphics[height=2cm]{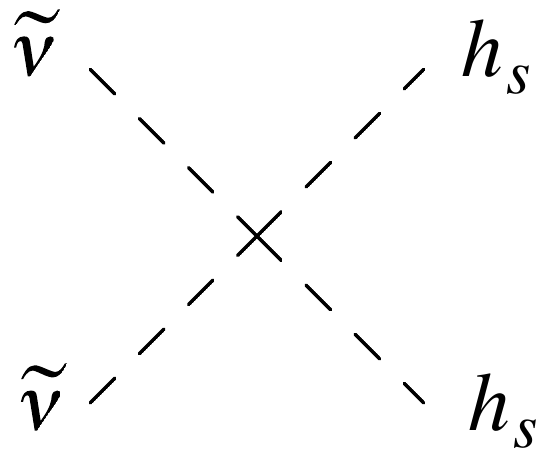}&
\raisebox{0.9cm}{	
	\bt{l}
		$\buildrel \sss{L}\over\sim C^2\epsilon_{sd}^2$\\
		$\buildrel \sss{R}\over\sim \lambda_N^2$
	\et
	}
\et
\caption{\label{fig:feynSnu}Tree level pair annihilation channels for the sneutrino into
$f\bar f$/$h_Sh_S$. Leading coefficients for the left-handed (L) and right-handed (R) case
are given. In the left-handed case the addition of a light singlet to the spectrum opens
up new channels but they are all suppressed by additional factors of $C^2$ or
$\epsilon_{sd}^2$ when compared to the originally dominant weak decay channel. In the
right-handed case the new $h_S$ final state channels are of the same order as the other decay
channels.}
\ec
\efig
Fig.~\ref{fig:feynSnu} shows the pair annihilation channels for the sneutrino LSP. In the
case of a left-handed sneutrino the channels decaying into $h_Sh_S$ are suppressed by
additional factors of $C^2$ and/or \esd. It is known that the left-handed sneutrino is a
poor LSP candidate if one wants to saturate the dark matter relic density limit
\cite{Falk:1994es}. This is caused by the fact that the dominant $\sim C^2$ weak decay
channel is always present for the left-handed sneutrino.

In the case of a right-handed sneutrino Fig.~\ref{fig:feynSnu} shows that the $h_Sh_S$
channels are all of the same order as the $f\bar f$ channels and collectively depend to
 leading order on $\lambda_N$ or $\lambda_N^2$. Thus, it is expected that the relic
density will be lower when compared to the CP-conserving case. The left plot in Fig.
\ref{fig:RDvsLamN} shows how the relic density for right-handed sneutrinos depends
strongly on $\lambda_N$, while the relic density for left-handed sneutrinos stays almost a
constant. The right plot in Fig. \ref{fig:RDvsLamN} shows the drop of the relic density
for the CP-violating case for right-handed sneutrino LSPs when compared to the
CP-conserving data. Although much of the set overlaps there is a clear systematic drop in
the mean relic density for the CP-violating set.

\bfig[ht]
\bc
\vspace{0.5cm}
\begin{tabular}{cccc}
\begin{sideways}\hspace{1.5cm}$\Omega_c h^2$\end{sideways}&
\includegraphics[width=0.4\textwidth]{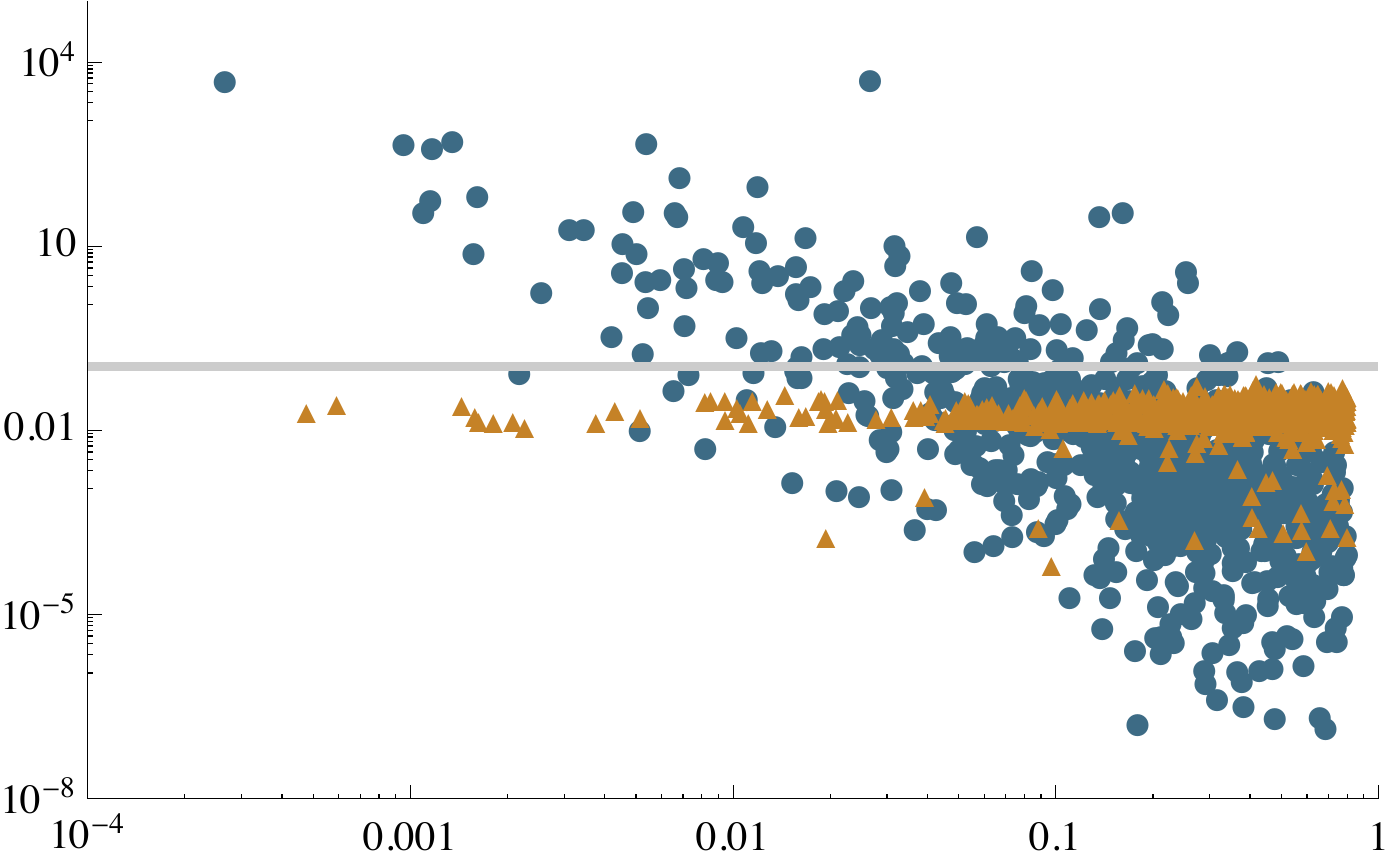}&
\begin{sideways}\hspace{1.2cm}$\log (\Omega_c h^2)$\end{sideways}&
\includegraphics[width=0.4\textwidth]{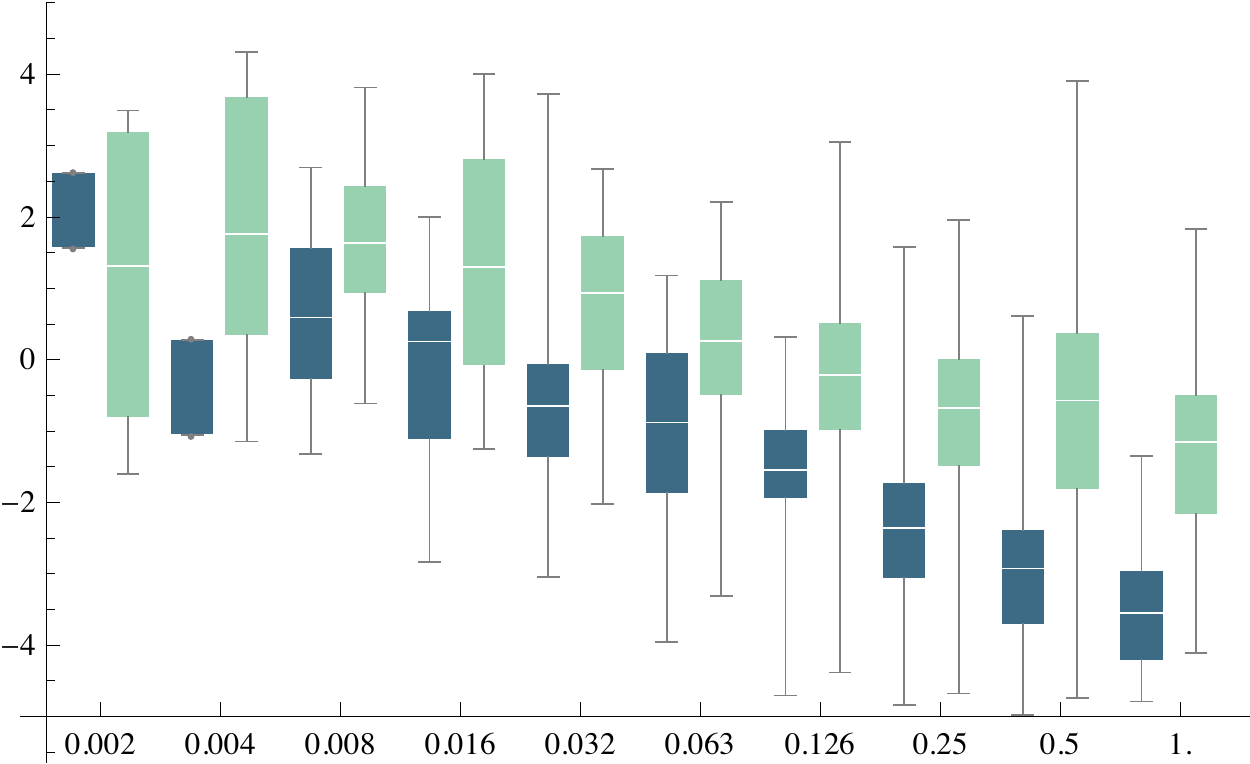}\\	
& \raisebox{0.1cm}[-0.1cm]{$\lambda_N$}
&& \raisebox{0.1cm}[-0.1cm]{$\lambda_N$}
\end{tabular}
\vspace{-0.3cm}
\caption{\label{fig:RDvsLamN}Left: The relic density against the trilinear coupling $\lambda_N$
for right-handed (blue circles) and left-handed (orange triangles) sneutrino LSPs. The
grey band indicates the current WMAP limits on the relic density. Right: A box-and-whisker diagram of the relic density against the trilinear coupling $\lambda_N$ comparing the distribution
of right-handed sneutrino LSPs in the CP-conserving (light
green) and CP-violating (blue) cases. Boxes are the 25 - 75 percentile range and whiskers
denote the complete range of the data.}
\ec
\efig

\bfig[ht]
\bc
\vspace{0.5cm}
\begin{tabular}{cccc}
\begin{sideways}\hspace{1.5cm}$\Omega_c h^2$\end{sideways}&
\includegraphics[width=0.4\textwidth]{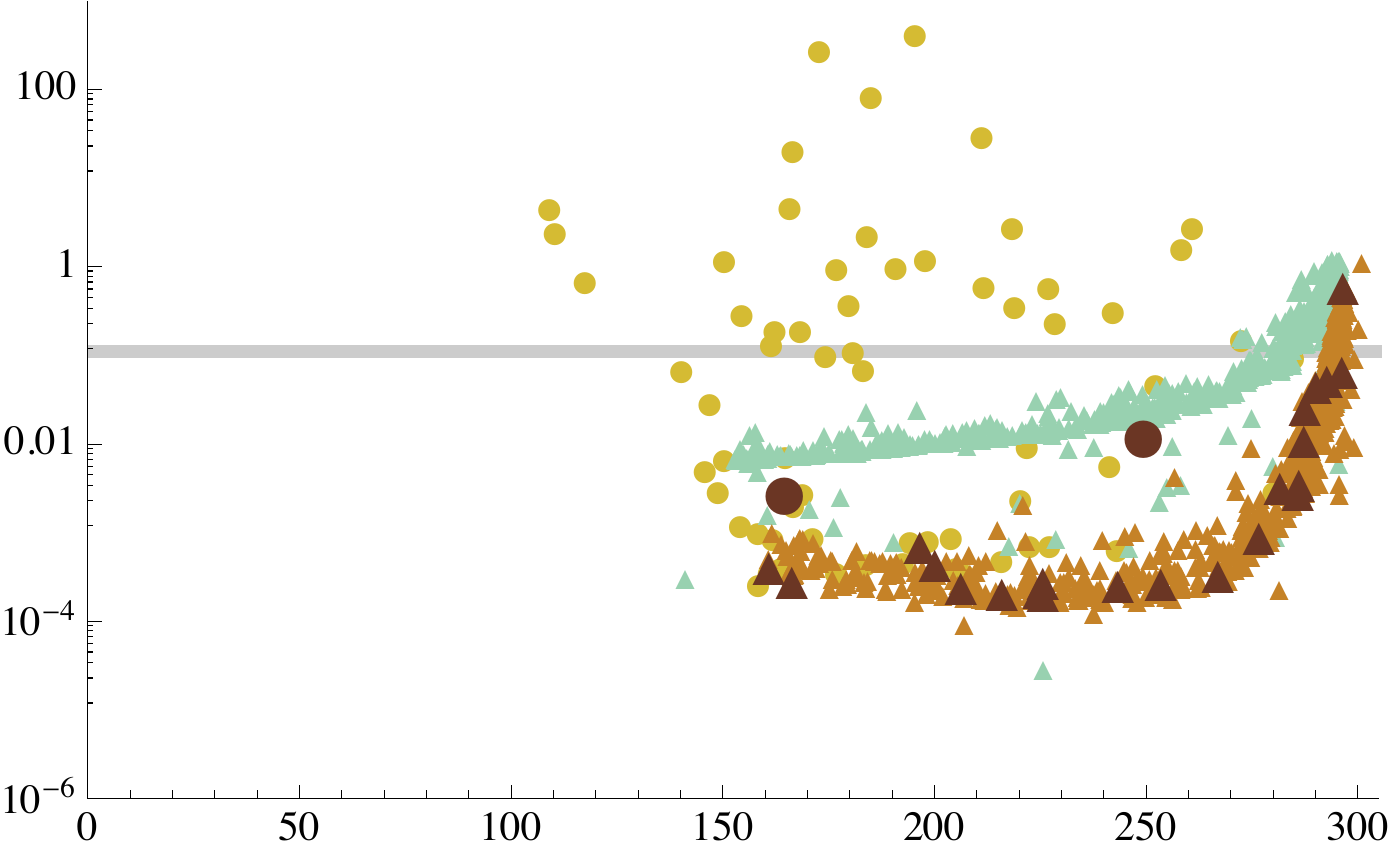}&
\begin{sideways}\hspace{1.5cm}$\Omega_c h^2$\end{sideways}&
\includegraphics[width=0.4\textwidth]{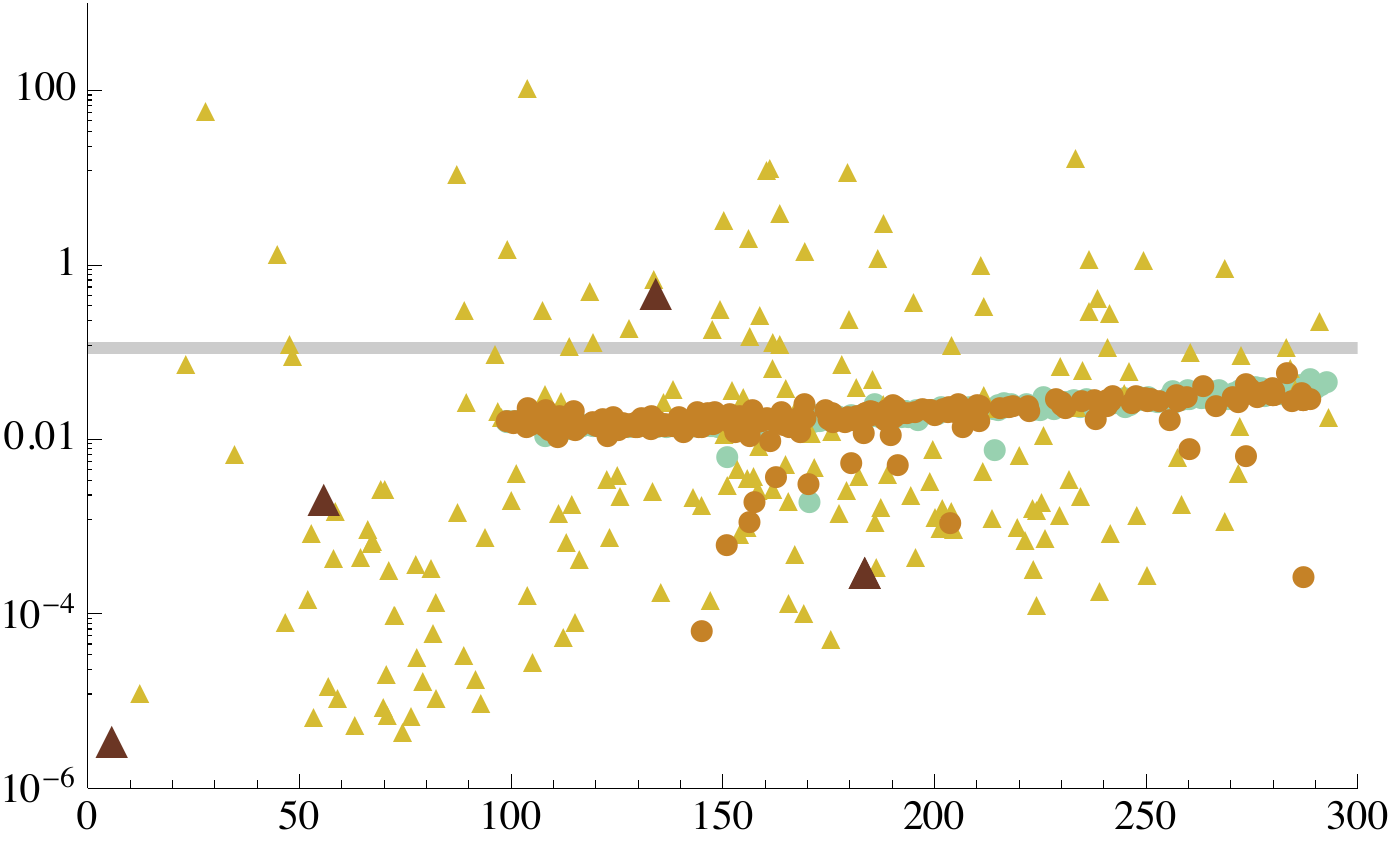}\\
& \raisebox{0.1cm}[-0.1cm]{$m_{\Neu 1}$}
&& \raisebox{0.1cm}[-0.1cm]{$m_{\tilde\nu}$}
\end{tabular}
\vspace{-0.3cm}
\caption{\label{fig:omadata1}The relic density against the LSP mass. Left: for neutralino LSPs
in the CP-conserving (light green triangle), CP-violating doublet dominated (orange
triangle) and CP-violating significant singlino admixture, i.e. $\ecs>0.1$, (yellow dot) cases.
Right: for sneutrino LSPs in the CP-violating left-handed (orange dot) and CP-violating
right-handed (yellow triangle) case. The CP-conserving left-handed (light green dot) set
overlaps completely with the CP-violating counterpart. In both plots all points satisfy
PDG constraints on the mass spectrum and vacuum stability, large brown triangles or
circles indicate points that pass all of the other experimental constraints we impose. The
grey band indicates the current WMAP limits on the relic density.}
\ec
\efig
In Fig.~\ref{fig:omadata1} all of the aforementioned effects are shown in unison. In the
plot on the left the fixed gaugino mass $M_1=300$ GeV causes the neutralino relic density
to spike at $m_{\sss LSP}=M_1$ due to the small bino annihilation process cross section.
We also see that for other values of $m_{\sss LSP}$ the neutralino has to be singlino
dominated (yellow dot) if we want to saturate the relic density limit and how in the case
of CP violation the relic density is significantly reduced compared to the CP conserving case.
 In the right plot we see that
left-handed sneutrino LSPs are poor candidates for saturating the relic density limit
regardless of CP violation, whereas for right-handed sneutrinos this is not a problem. The
scattered very low relic density points for the left-handed sneutrinos are due to the
presence of efficient slepton co-annihilations. Both plots also show how few points pass
all of the experimental constraints we impose. We find that the most restrictive
constraint is the electron EDM, e.g., in Fig.~\ref{fig:omadata1} removing all other CP, B
physics, and rare decay constraints only yields a single additional ``allowed" point.

In general we find that for constrained phases the effect on the relic density from
variations in the spectrum of the model is much smaller than the effect introduced by the
appearance of a light singlet scalar providing new annihilation channels for the LSP.
Depending on the LSP and its composition these channels are governed by the Lagrangian
parameters $\kappa$ and $\lambda_N$.

\section{Direct dark matter searches}
\label{sec:direct-detection}

There are currently several ongoing searches of weakly interacting massive particles
(WIMPs) in direct detection experiments. The goal of the experiments is to measure the
WIMP--nucleus interaction in the detector material. WIMPs scatter when they interact with
 nuclei, and the recoil energy can then be measured. As yet, the XENON100 experiment
puts the tightest limits on the WIMP-nucleon scattering cross section
\cite{Aprile:2011dd}. To compare the calculated proton/neutron cross sections with the
experimental limits, we use a normalized cross section for a point-like nucleus
\cite{Belanger:2010cd}:
\begin{equation}
  \label{eq:SIxs}
  \sigma_\text{SI} = \frac{(Z \sqrt{\sigma_{p}} +
    (A-Z)\sqrt{\sigma_{n}})^2}{A^2},
\end{equation}
$Z$ and $A$ being the atomic and mass number of the target element\footnote{For Xenon:
$A=131,\ Z=54$.} and $\sigma_{n,p}$ the spin-independent cross-sections for neutron and
proton target, respectively. Because there are large uncertainties in the local density of
dark matter and in the nuclear matrix elements that enter the computation of
$\sigma_\text{SI}$ \cite{Ellis:2005mb,Belanger:2008sj}, the direct detection limits are
only indicative.

\begin{figure}[htb]
\centering
\includegraphics[width=0.6\textwidth]{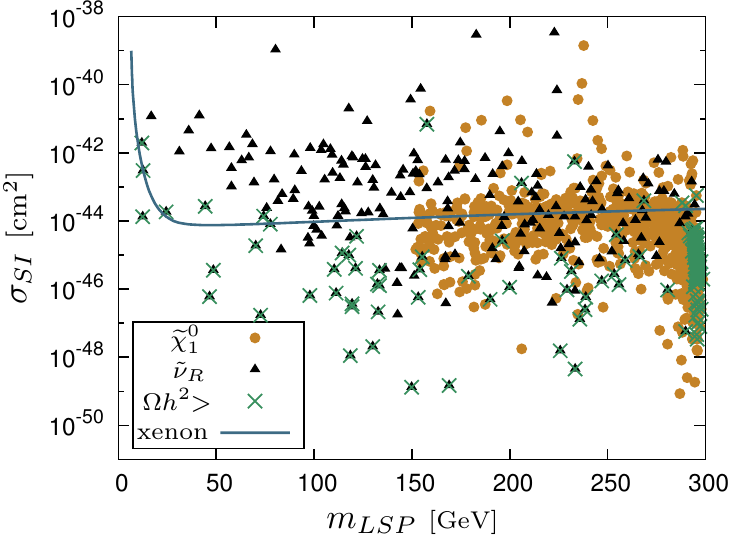}
\caption{Spin-independent WIMP--nucleon cross section for CPV
  points. Points with triangular shape (black) have the right-handed
  sneutrino as an LSP, while the dots (orange) have neutralino LSP. The
  (green) cross above a point corresponds to a relic density above
  the WMAP upper limit, hence an excluded point. The
  solid line represents the XENON100 experimental limit.\label{fig:directdet}}
\end{figure}

Fig.~\ref{fig:directdet} shows the LSP mass versus the spin independent cross section for
a set of CP-violating points, as well as the XENON100 limit.
We see that the XENON100 limit actually serves as a constraint, since a relatively large
portion of the parameter points are ruled out by the direct detection search. A roughly equal
number of points are still allowed by the XENON100 limit. The WMAP upper limit constraint
removes most of the sneutrino LSP points allowed by the XENON100 limit. However, some
points still evade both dark matter constraints. A large number of neutralino LSP points
are still allowed in the low neutralino mass region. In ref.~\cite{Perelstein:2012qg} it
was found that in the CP-conserving NMSSM the direct detection limits for the neutralino
LSP still allow a large number of points with low fine tuning, especially for a large
$\lambda$ parameter.

\section{Signatures at the colliders}
\label{sec:sign-at-coll}

In this section we focus on the possible signals from the supersymmetric particle spectrum
realized by the model respecting all relevant low energy constraints and generating
an acceptable amount of relic density. Notwithstanding the fact that we have already
restricted the parameter space of the model significantly by various constraints, recent
observation of a $\sim 125-126$ GeV Higgs like boson at LHC by both the CMS
\cite{CMS:2012gu} and ATLAS \cite{Atlas:2012gk} Collaborations further helps in narrowing
down the allowed sparticle spectrum.

As discussed in Sec.~\ref{sec:cp-violation} in the model used here with SCPV a light scalar 
state $h_1\equiv h_S$ exists and thus the branching ratio of the decay of the $\sim 125$ 
GeV Higgs, $h_2$, into a pair $h_S$ is of particular interest. A limit on $\mathrm{BR} 
(h_2\to h_Sh_S)$ could be useful in further constraining this type model. We find that 
{\em a priori} requiring $m_{h_S}>m_{h_2}/2$ is problematic since a larger mass for the 
singlet type Higgs implies larger CP violation and thus the limits on the EDMs exclude 
these points. Instead we extract the $hhh$ coupling as
\bea
C_{ijk}^{hhh} &=& i \sum_{r,s,t}^{6,6,6} 
\mathcal{O}_{ir} \mathcal{O}_{js} \mathcal{O}_{kt}
\frac{\delta}{\delta \phi_r}\frac{\delta}{\delta \phi_s}\frac{\delta}{\delta \phi_t}  
\mathcal{L}_{int}\genfrac{}{}{0pt}{}{}{|_{\rm vev}},\\
\phi&\in&\{\mathrm{Re}H_1^0,\mathrm{Re}H_2^0,\mathrm{Re}S,\mathrm{Im}H_1^0,\mathrm{Im}H_2^0,\mathrm{Im}S\}
\eea
with $\mathcal{O}$ the Higgs mixing matrix and require $|C^{hhh}_{211}|<1$GeV.
From the points that clear this limit we have selected a few which also clear the EDM,
B-physics, and rare decay constraints and are favored by relic density and analyze
their signal at LHC. We have listed the benchmark points (BP) in Table \ref{tab:benchpts}.
Note that LHC has already put some strong limits on the squark and gluino masses
\cite{CMS:2012jx,Collaboration:2012rz}, albeit these constraints are model dependent and
also depend on the way the heavy particles decay. However, for our analysis, we have made
a consistent choice of parameters where the strongly interacting sector of the sparticle
spectrum remains above a TeV and is, therefore, unconstrained from LHC data. This implies
that our analysis will focus on the weakly interacting sector of the spectrum
\textit{viz.}, sleptons, gauginos/higgsinos and the Higgs.

Our choice for parameter points leads to a right-handed sneutrino LSP or a neutralino LSP.
For \texttt{BP6} and  \texttt{BP7} the LSP comes out to be a right-handed sneutrino
with a mass of $\sim 6$ GeV and $\sim 184$ GeV, respectively. All the other benchmark
points have a neutralino as an LSP. We focus on each benchmark point individually and
highlight the most likely signals for the different parameter choices at LHC in our model.

\begin{table}[ht]
\bc
\begin{minipage}[b]{1\linewidth}\centering
\small
\bt{|c|c|c|c|c|c|c|c|c|}
\hline
        & \texttt{BP1} &\texttt{BP2} &\texttt{BP3} &\texttt{BP4} &\texttt{BP5} &\texttt{BP6} &\texttt{BP7} \\ \hline
$\tb$       &45.7    & 12.4   & 33.9   & 26.9  & 30.1  & 43.5   & 44.9   \\
$\lambda$   & 0.114  & 0.355  & 0.363  & 0.347 & 0.430 & 0.259  & 0.341  \\
$\kappa$    & 0.038  & 0.63   & 0.231  & 0.294 & 0.357 & 0.174  & 0.292  \\
$v_S$ (GeV) & 4277.2 & 1160.2 & 964.8  & 914.8 & 487.5 & 1881.9 & 816.7  \\ \hline
$\lambda_N$ & 0.721  &0.008   & 0.032  & 0.498 & 0.269 & 0.035  & 0.750  \\
$A_{\lambda_N}$ (GeV) 
            & 337.0  & 25.2   & -668.9 & 319.7 & 365.3 & -135.5 & -975.5 \\
$M_N$ (GeV) & 447.7  & 449.8  & 494.7  & 401.1 & 341.5 & 207.6  & 135.7  \\
$M_{L,E}$ (GeV) 
            & 307.1  & 419.5  & 432.0  & 431.9 & 472.7 & 320.3  & 377.0  \\ \hline
$\delta_S$  & 3.228  & 0.156  & 3.142  & 3.203 & 3.213 & 3.199  & 3.173  \\
$\delta_2$  & 0.111  & 0.034  & 0.010  & 0.249 & 3.037 & 0.142  & 0.173  \\
$\xi$ (GeV) & 118.1  & -991.3 & 522.9  & 318.9 & 253.0 & 109.8  & 300.6  \\
\hline
LSP-type    & $\chi_0$ & $\chi_0$ & $\chi_0$ & $\chi_0$ & $\chi_0$ & $\tilde\nu_R$ & $\tilde\nu_R$ \\
$m_{\sss LSP}$ (GeV) 
            & 296.3  & 290.9  & 286.3  & 276.7 & 194.6 & 5.7   & 183.7\\
$\Omega_ch^2$ 
            & 0.062  & 0.098  & $8.3\cdot 10^{-3}$ 
& $8.3\cdot 10^{-4}$ & $6.5\cdot 10^{-4}$ & $3.3\cdot 10^{-6}$ & $2.9\cdot 10^{-4}$ \\
\hline
\et
\end{minipage}
\caption{\label{tab:benchpts} Benchmark points for studying the signals of the
supersymmetric particles at LHC. Note that  $\mu=\lambda  v_S $ and other parameters are
the same as shown in Table \ref{tab:parCommon}. }
\ec
\end{table}

\pagebreak
\begin{itemize}

\item \textbf{BP1}

This point gives a neutralino LSP with mass $\sim 296$ GeV while the lightest sneutrino is
nearly degenerate at $\sim 300$ GeV. The chargino $\chi_1^+$ is much heavier (470 GeV)
than the second neutralino $\chi^0_2$ (322 GeV). Although this point is allowed by all
experimental data it does not give any significant cross sections at LHC in any channel. 
The chargino pair production which is the largest is about 8 fb at LHC with $\sqrt{s}=14$
TeV. The dominant decay for the chargino is to charged leptons and lighter sneutrinos
which then decay invisibly. So the final state which could be observed at LHC is
$\ell^+_i\ell^-_j\slashed{E}_T$ which would require substantial luminosity to separate it
from the large SM background coming from the $W^+W^-$ production. The point also gives a
very light scalar of $\sim 2.7$ GeV which is dominantly singlet with very suppressed
couplings to all SM particles, such that it is not constrained by any experimental data.

\item \textbf{BP2}

For this point we have a relatively small $\tan\beta$ and $\lambda_N$ as compared to all
other points. The LSP is the lightest neutralino which has a mass of $\sim 291$ GeV. The
small value of $\lambda_N$ leads to a very light right-handed neutrino of mass $\sim 19$
GeV. This state has a very small mixing with the left-handed neutrino and thus is not
constrained by the measured $Z$ width or other low energy data. The strongest signal for
this point at LHC is through the production of first chargino and second neutralino 
($\chi^+_1 \chi^0_2$) whose mass is about 400 GeV with a cross section of $\sim 5$ fb for
the current run at LHC with $\sqrt{s}=8$ TeV while the next significant production cross
section is for the pair production of charginos ($\chi^+_1 \chi^-_1$) with a cross section
of  $\sim 4$ fb. All other production modes are suppressed. The cross sections are
increased by a factor of $\sim 3-4$ at LHC with  $\sqrt{s}=14$ TeV while other modes also
become accessible with high luminosity. However, we note that the mass difference of these
states with the LSP is only about 100 GeV. Both the $\chi_1^+$ and $\chi^0_2$ decay to the
LSP and a weak gauge boson with 100\% branching probability. Thus the pair production of
charginos will lead to a final state of $W^+W^-\slashed{E}_T$ where both $W$'s will have
small transverse momenta. The $\chi^+_1 \chi^0_2$ production leads to final states with
$W^+Z \slashed{E}_T$ with no suppression in branching ratios. This leads to interesting 
signal of a Z peak, an associated charged lepton and large missing energy. The SM 
background will be mostly from $WZ$ and triple gauge boson production where the large 
$WZ$ production can be suppressed by demanding large $\slashed{E}_T$.

\item \textbf{BP3}

This point gives a neutralino LSP with a mass of  $\sim 286$ GeV with next lightest
supersymmetric particles being $\chi^0_2$ (338 GeV) and $\chi_1^+$ (341 GeV). This point
also gives us a light right-handed neutrino with mass $61$ GeV and an ultralight scalar
with mass 300 MeV. The largest production cross section in this case is again $\chi^+_1
\chi^-_1$ and $\chi_1^+\chi^0_2$ but the $\chi_1^+$  decays via off-shell weak
gauge boson ($W^+$) and off-shell sleptons/sneutrinos leading to final state particles
with small transverse momenta. The $\chi^0_2$ decays to an ultralight singlet scalar and
the LSP with 100\% probability. The signal production cross sections are about $\sim 7.5$
fb and $\sim 3$ fb, respectively, at LHC with $\sqrt{s}=8$ TeV, and a factor $\sim 3-4$
larger at LHC with $\sqrt{s}=14$ TeV. As the final decay products do not carry large
transverse momenta, isolating the signal from the background will be very challenging.

\item \textbf{BP4}

This point has a very similar spectrum to \texttt{BP3} except for a much heavier
right-handed neutrino with a mass of $\sim 910$ GeV as $\lambda_N$ is much larger. The
lightest scalar is the dominantly singlet component of the Higgs with mass $\sim 17$ GeV.
The neutralino LSP has a mass of $\sim 277$ GeV. The next to lightest supersymmetric
particle (NLSP) is the $\chi_1^+$ (311 GeV) while  $\chi^0_2$ (316 GeV) also is close and
nearly degenerate. The mass difference again forces the decay similar to that for 
\texttt{BP3} with final states with small transverse momenta. The largest cross section in
this case is for the pair production of charginos ($\chi^+_1 \chi^-_1$) which is about 13
fb at LHC with $\sqrt{s}=8$ TeV which increases to $\sim 40$ fb for $\sqrt{s}=14$ TeV.

\item \textbf{BP5} 

This point corresponds to a much lighter supersymmetric spectrum with a neutralino LSP of
mass $\sim 195$ GeV. The NLSP is  $\chi^0_2$ (199 GeV) with $\chi_1^+$ (204.5 GeV) the
next lightest. It also has a 262 GeV right-handed neutrino. The light spectrum leads to
much larger production cross sections, and the dominant production channels in this case
are $\chi^+_1 \chi^-_1, \chi^+_1 \chi^0_1$ and $\chi_1^+\chi^0_2$ with production
cross sections of around 74 fb, 86 fb and 82 fb, respectively, at LHC with $\sqrt{s}=8$
TeV. However, the mass difference of $\chi^+_1$ and $\chi^0_2$ with $\chi^0_1$ is less
than 10 GeV, and thus the transverse energy of the decay products is small. It would be
quite difficult to trigger on such soft leptons and jets that result from the 3-body
decays of  $\chi^+_1$ and $\chi^0_2$. A detailed analysis would be required to isolate
signals which would be relevant to highlight for LHC searches.

\item \textbf{BP6}

Unlike the other points, we get a right sneutrino LSP for this parameter choice with a
mass of $\sim 6$ GeV. Pair production of the LSP has the largest cross section but will
pass through the detector undetected. The other significant cross sections ($\chi^+_1
\chi^-_1, \chi_1^+\chi^0_2$) are relevant only at LHC with $\sqrt{s}=14$ TeV with
production cross sections of  less than 8 fb. The decay of the  $\chi_1^+$ and $\chi^0_2$
is however different from that of \texttt{BP2} with suppressed probabilities in similar
channels and therefore the signal gets suppressed further. Much like \texttt{BP1} we get a
light scalar with a mass of  $\sim 2.5$ GeV. We also get a light right-handed neutrino of
mass 130 GeV which can lead to interesting signals if there is significant mixing with the
left-handed neutrinos.

\item \textbf{BP7}

Our final benchmark point gives us a right sneutrino LSP of mass
$\sim 184$ GeV. This point also gives a light scalar Higgs of mass $\sim 8.4$ GeV in the
mass spectrum. The $\chi^0_1$ (254 GeV) is the NLSP while the right-handed neutrino mass
is above 1 TeV. The largest production cross section in this case is for $\chi^+_1
\chi^-_1$ where the chargino mass is $\sim 273$ GeV, which at LHC with $\sqrt{s}=8$ TeV is
about 22 fb and jumps up to $\sim 65$ fb for $\sqrt{s}=14$ TeV. The chargino decays to a
$\tau$ and LSP with 60\% branching probability while it decays 20\% of the time to
electron/muon and LSP.  This leads to final states of two charged leptons with large
missing energy. Other relevant channels are from production of  $\chi^0_1\chi^0_2,
\chi^+_1\chi^0_1$ and  $\chi^+_1\chi^0_2$ where the  $\chi^0_1$ decays invisibly while the
 $\chi^0_2$ decays to the light scalar Higgs and  $\chi^0_1$ with 100\% probability.

\end{itemize} 

Although we discuss some possible signals at LHC for our model with different choices of
the parameter space a more detailed analysis of both signal and background is needed which
we leave for future work. The CP-violating effects due to non-zero phases can affect
several observables reconstructed from the cascade decays of the charginos, neutralinos or
the scalars in the model and have been highlighted in various studies
\cite{MoortgatPick:2010wp,Kittel:2011sq,Bornhauser:2011ab,Berger:2012gh} and we aim to
apply these studies in our future analysis of the model at LHC in our next publication
\cite{future_studies}.

\section{Summary and discussion}
\label{summary}

We have studied the viability of the right-handed sneutrino and neutralino dark matter in
the NMSSM, and the effect of CP violation on the dark matter relic density and the direct
detection potential. The studied model contains, in addition to the MSSM fields, two
singlet superfields and interaction terms. This extends the Higgs sector, adds the
right-handed neutrino field to the model, and allows the right-handed sneutrino to
become a thermally produced dark matter candidate. We took into account the \Bsg, \Btn, and \Bmumu
bounds, the discovery of the Higgs and sparticle searches, and constrained the model
parameters by calculating the dark matter relic density, including co-annihilations. We
calculated also the direct detection rates for the nucleon--DM spin-independent cross
section, and compared that to the XENON100 limit. We have found that for constrained
CP-violating phases the effect on the relic density from variations in the spectrum of the
model is much smaller than the effect introduced by the appearance of a light singlet
scalar providing new annihilation channels for the LSP. In particular for neutralinos and
right-handed sneutrinos these new channels lower the relic density. In some cases, the
left-handed sneutrinos co-annihilate very efficiently with other sleptons resulting in
very low relic density. The spin-independent cross section can be used to constrain the
models with a light sneutrino LSP, as for these models the direct detection limit tends to
be complementary to the relic density constraint. The neutralino dark matter, however,
remains rather unconstrained by the direct detection limits. We have also used our
analysis to identify a set of benchmark points which satisfy all constraints and favor DM
data. We have used these benchmark points to highlight the most important signals for the
model that can be observed at the current and future run of the LHC.

\section{Acknowledgments}
KH, LL, and TR acknowledge support from the Academy of Finland (Project No. 137960).  The
work of JL is supported by the Foundation for Fundamental Research of Matter (FOM),
program 104 ``Theoretical Particle Physics in the Era of the LHC".  LL thanks Magnus
Ehrnrooth Foundation for financial support. SKR is supported in part by the United States
Department of Energy, Grant Number DE-FG02-04ER41306.

\bibliographystyle{JHEP}
\bibliography{scdmbib.bib}

\end{document}